# Enhancing the three-dimensional electronic structure in 1111-type iron arsenide superconductors by H-substitution

Yoshinori Muraba[1], Satoru Matsuishi[2], Hideo Hosono[1,2,3*]

[1]*Materials and Structures Laboratory, Tokyo Institute of Technology, 4259 Nagatsuta-cho, Midori-ku, Yokohama 226-8503, Japan*

[2]*Materials Research Center for Element Strategy, Tokyo Institute of Technology, 4259 Nagatsuta-cho, Midori-ku, Yokohama 226-8503, Japan*

[3]*Frontier Research Center, Tokyo Institute of Technology, 4259 Nagatsuta-cho, Midori-ku, Yokohama 226-8503, Japan*

*Corresponding author, Email: hosono@lucid.msl.titech.ac.jp

Phone: +81-(0)45-924-5359 Fax: +81-(0)45-924-5196

**ABSTRACT**

The 1111-type iron arsenide hydride CaFe$_{1-x}$Co$_x$AsH was synthesized by high-pressure solid-state reaction, and its electronic structure and superconducting properties were investigated. Bulk superconductivity was observed at $x$ = 0.09–0.26. A maximum superconducting critical temperature ($T_c^{max}$) of 23 K was observed at $x$ = 0.09. These values are in agreement with those of CaFe$_{1-x}$Co$_x$AsF. The calculated Fermi surface of CaFeAsH has a small three-dimensional (3D) hole pocket around the Γ point. This is a result of weak covalent bonding between the As 4$p$ and H 1$s$ orbitals. No such covalency exists in CaFeAsF, because the energy level of the F 2$p$ orbital is sufficiently deep to inhibit overlap with the As 4$p$ orbital. The similar superconductivities of CaFe$_{1-x}$Co$_x$AsH and CaFe$_{1-x}$Co$_x$AsF are explained on the nesting scenario. The small 3D hole pocket of CaFe$_{1-x}$Co$_x$AsH does not significantly contribute to electron excitation. These findings encourage exploration of hydrogen-containing 1111-type iron-based materials with lower anisotropies and higher $T_c$ applicable to superconducting wires and tapes.

I. INTRODUCTION

Since the discovery of superconductivity in LaFeAsO$_{1-x}$F$_x$ (critical temperature ($T_c$) = 26 K),[1] the physical properties of new families of iron pnictide superconductors have been extensively investigated.[2–7] $Ln$FeAsO ($Ln$ = lanthanide)[1,8–11] and $Ae$FeAsF ($Ae$ = alkali earth),[12,13] or the so-called 1111-type oxyarsenides and fluoroarsenides, respectively, are typical such families. They possess a ZrCuSiAs-type crystal structure, composed of alternately stacked FeAs anti-fluorite-type conducting layers and $Ln$O/$Ae$F fluorite-type insulating layers. Stoichiometric 1111-type families undergo structural and magnetic transition with decreasing temperature. Superconductivity arises when these transitions are suppressed, by carrier doping of the FeAs layers. Carrier doping modes for these 1111-type families are classified as "direct" and "indirect" doping, according to their substitution sites. The former involves elemental substitutions in FeAs layers, such as cobalt or nickel substitution at iron sites.[14–17] The latter involves aliovalent ion substitutions in $Ln$O/$Ae$F layers, such as substitution at oxygen sites with fluorine.

We recently reported the high-pressure synthesis of the hydrogen-substituted 1111-type compounds CaFeAsF$_{1-x}$H$_x$ ($x$ = 0.0–1.0)[18] and $Ln$FeAsO$_{1-x}$H$_x$ ($x$ = 0.0–0.5).[18–20] Hydrogen is incorporated as H$^-$ at F$^-$ or O$^{2-}$ sites, in the blocking layers of CaFeAsF or $Ln$FeAsO, respectively. The high solubility of hydrogen ($x$ < 0.5) in 1111-type oxyarsenides results in a wide superconducting dome of 0.05 < $x$ < 0.4–0.5. The valence state and ionic radius of hydrogen are close to those of fluorine. However, their solubility toward oxygen and their pressure dependence of $T_c$ are rather different.[21] This implies that the large spatial spread and softness of hydride electrons lead to different chemical bonding states in 1111-type iron pnictides and the fluoride-1111 material.

In this paper, we report the superconductivity of 1111-type CaFeAsH, which is induced by Co-substitution at the Fe site. The maximum $T_c$ = 23 K and the extent of the superconductivity (SC) region in the $x$-$T$ diagram are similar to those of Co-substituted CaFeAsF. The effect of hydrogen on 1111-type iron pnictides is discussed, by comparing the electronic structures of $CaFe_{1-x}Co_xAsH$ and $CaFe_{1-x}Co_xAsF$.

## II. EXPERIMENTAL

$CaFe_{1-x}Co_xAsH$ was synthesized by the solid-state reaction of CaAs, $CaH_2$, $Fe_2As$ and $Co_2As$, using a belt-type high-pressure anvil cell. The specific reaction was:

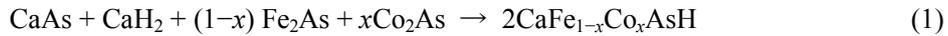

$$CaAs + CaH_2 + (1-x)\,Fe_2As + xCo_2As \rightarrow 2CaFe_{1-x}Co_xAsH \qquad (1)$$

$CaH_2$, CaAs, $Fe_2As$ and $Co_2As$ were prepared from their respective metals. $CaH_2$ was synthesized by heating metallic Ca in a $H_2$ atmosphere. All starting materials and precursors were prepared in a glove box under a purified Ar atmosphere ($H_2O$ and $O_2$ < 1 ppm). The starting material mixture was placed in a BN capsule, with a mixture of $Ca(OH)_2$ and $NaBH_4$ acting as an excess hydrogen source.[22] The mixture was heated at 1273 K under 2 GPa for 30 min. The crystal phase and structure were identified by powder X-ray diffraction (XRD), using Mo $K_\alpha$ radiation at room temperature. The sample was first ground into a fine powder, and placed in a glass capillary ($\varphi$ 0.5 mm). XRD measurements were recorded in transmission mode, to reduce the effects of preferential crystallite orientation. Rietveld refinement of XRD patterns was performed using TOPAS software.[23] The chemical composition was determined by an electron-probe micro analyzer (EPMA; JEOL model JXA-8530F), equipped with a field-emission-type electron gun and wavelength dispersive X-ray detectors. Micron-scale compositions within the main phase were probed at ten individual focal points, and the results were then averaged. The temperature dependence of the DC electrical resistivity ($\rho$) at 2–200 K was measured using the conventional four-probe method, using Ag paste as the electrodes. Magnetic susceptibility ($\chi$) was measured with a vibrating sample magnetometer (Quantum Design). $\chi$ was measured on bulk and powdered samples and the data of the powder were adopted. Non-spin-polarized density functional theory (DFT) calculations with Perdew-Burke-Ernzerhof functional[24,25] and the projected augmented plane-wave method[26] were implemented in the Vienna ab initio Simulation Program (VASP) code.[27] A conventional cell containing two chemical formula units, and structural parameters obtained by the interpolation of experimental values, were used. The effect of partial Co-substitution at the Fe site was taken into account through the virtual crystal approximation. The plane-wave basis-set cutoff was 600 eV. For Brillouin-zone integrations to calculate the total energy and DOS, 20×20×10 Monkhorst-Pack grids of $k$ points were used. The projected density of states (PDOS) was obtained by decomposing the charge density over the atom-centered spherical harmonics with the same Wigner-Seitz radius $r = (3V_{cell}/4\pi N)^{1/3}$, where $V_{cell}$ and $N$ are the unit-cell volume and number of atoms per unit cell, respectively.

## III. RESULTS

**Crystal structure**

Figure 1(a) shows the powder XRD pattern of $CaFe_{1-x}Co_xAsH$, with $x$ value of 0.12. Except for minor peaks arising from Fe metal and unknown impurities, most peaks can be indexed to a ZrCuSiAs-type structure of $P4/nmm$ symmetry, as shown inset in Fig. 1(a). Figure 1(b) shows the analyzed $x$ value in $CaFe_{1-x}Co_xAsH$, as a function of ($x_{nom}$) in the starting mixture. The analyzed $x$ value is proportional to $x_{nom}$, and its deviation from $x_{nom}$ indicates the segregation of a Co-rich impurity phase. The EPMA analysis indicates the existence of an impurity of composition $CaFe_{1.85\pm0.12}Co_{1.03\pm0.11}As_{1.82\pm0.05}$, in addition to Fe metal. The volume fraction of Co-rich impurity phase estimated by analyzing the EPMA mapping images (200 μm × 200 μm, 400 × 400 points) linearly increases with increasing $x$ and reaches up to 27 % at $x = 0.308$. Figure 1(c) and (d) show variations in the $CaFe_{1-x}Co_xAsH$ lattice parameters $a$ and $c$, respectively, as a function of analyzed $x$. These parameters for $CaFe_{1-x}Co_xAsF$ are also shown for comparison.[28] The $CaFe_{1-x}Co_xAsH$ $a$-axis length increases with increasing Co-substitution, whereas the *c*-axis length decreases. The *a*-axis dimensions of $CaFe_{1-x}Co_xAsH$ are comparable to those of $CaFe_{1-x}Co_xAsF$, whereas the *c*-axis lengths are ≈35 pm shorter than those of $CaFe_{1-x}Co_xAsF$. Figure 1(e), (f) and (g) show the As distance from the Fe plane ($h_{As}$), the Ca distance from the H/F plane ($h_{Ca}$), and the distance between the Ca and As planes ($d_{Ca-As}$), respectively. While $h_{Ca}$ and $d_{Ca-As}$ remain largely constant, $h_{As}$ decreases with increasing *x*. This indicates that the monotonic decrease of *c*-axis length with increasing *x* originates from the decrease in $h_{As}$. The different $h_{As}$ (< 2 pm), $h_{Ca}$ (≈ 7 pm) and $d_{Ca-As}$ (≈ 12 pm) values of $CaFe_{1-x}Co_xAsH$ and $CaFe_{1-x}Co_xAsF$ are independent of Co-substitution. This also supports the conclusion that the different *c*-axis lengths originate from the decrease in $d_{Ca-As}$.

**Superconducting Properties**

Figure 2(a) and (b) show the temperature dependence of the electrical resistivity ($\rho$) for $CaFe_{1-x}Co_xAsH$, with *x* = 0.04–0.12 and 0.17–0.31, respectively. The anomaly in the $\rho$-*T* curves due to structural or magnetic transitions is observed at $T_{anom}$ ≈ 80 K for *x* = 0.02. $T_{anom}$ decreases to ≈ 55 K at *x* = 0.04. These samples exhibit a small $\rho$ decrease at ≈ 20 K, but zero resistivity is not observed. Figures 2(c) and (d) show the temperature dependence of the magnetic susceptibility ($\chi$) for $CaFe_{1-x}Co_xAsH$, with *x* = 0.04–0.12 and 0.17–0.31, respectively. The $4\pi\chi$ value of near 0 at *x* = 0.02 and 0.04 indicates that the $\rho$ decrease of these samples arises from the local inhomogeneity of Co. As *x* increases ≥ 0.07, zero resistivity is attained and the onset $T_c$ reaches a maximum ($T_c^{max}$) of 23 K at

$x = 0.07$. Further Co-doping causes a monotonic decrease in $T_c$. Superconductivity is eventually completely suppressed at $x = 0.31$. Figure 2(e) shows the shielding volume fraction (SVF) evaluated from the gradient of the $M$ vs. magnetic field ($H$) curve at 2 K and the volume fraction of the Co-rich impurity phase estimated by EPMA. The SVF was evaluated by utilizing the volume fraction of contained phases in samples estimated by EPMA on the assumption that the density of the Co-rich impurity phase was the same as that of $CaFe_2As_2$ whose constituent element and composition is very close to the Co-rich impurity phase. The SVF value of $> 26\%$ for samples of $0.09 \leq x \leq 0.26$ indicates bulk superconductivity. Here, we mention that the possibility of superconductivity derived from the Co-rich impurity. Since there are no step on the $4\pi\chi$-$T$ curves in the superconducting region, superconductivity observed here is probably caused by a unity phase unless superconducting domes of the others phases accidentally coincide. Moreover, the volume fractions of the Co-impurity phase are smaller than SVF values in the range of $0.09 \leq x \leq 0.26$. Consequently, it is concluded that observed superconductivity is caused by $CaFe_{1-x}Co_xAsH$. Figure 2(f) shows $T_{anom}$ and $T_c^{onset}$ values from the $\rho$-$T$ curves, as a function of $x$. These values for $CaFe_{1-x}Co_xAsF$ are also shown for comparisoin.[12] $T_c$-dome values for $CaFe_{1-x}Co_xAsH$ are in the range of $x = 0.09$–$0.26$, with $T_c^{max} = 23$ K. The $T_c^{max}$ and the extent of the SC region are comparable to those of $CaFe_{1-x}Co_xAsF$. This implies that the superconducting properties are not significantly changed by replacing the blocking layer anion from $F^-$ to $H^-$.

**Electronic structures**

Figure 3(a) compares the calculated DOS of CaFeAsF and CaFeAsH. Energy bands located around the Fermi level ($E_F$) are mainly composed of Fe $3d$ states. The DOS of these two compounds are comparable at $-2 < E < 2$ eV. Below the Fe $3d$ bands, the energy bands are mainly composed of As $4p$ states, which are located at $-6 < E < -2$ eV. In CaFeAsF, fluorine $2p$ states form further bands at $-7.5 < E < -5.5$ eV, without admixing with the orbitals of other atoms. On the other hand, in CaFeAsH, hydrogen $1s$ states energetically overlap well with As $4p$ states. Ca orbitals contribute at $-2 < E < -5$ eV, which suggests that they admix with the H1s orbital. Figure 3(b) shows band dispersion and the contribution of Fe $3d$ orbitals. In CaFeAsF, there are three hole pockets around the Γ point, and two electron pockets around the M-point. The dispersion and orbital contribution of the electron-pockets and two of the three hole-pockets are comparable with those of CaFeAsH. However, one hole-like band in CaFeAsH crosses the $E_F$ in the Γ-Z path, as shown by the red line in Fig. 3(b) (which corresponds to the crystallographic $c$-axis). As a consequence, the dominant orbital nature in the hole pocket changes from Fe $3d_{yz/zx}$ in CaFeAsF, to Fe $3d_{x^2-y^2}$ and $3d_{z^2}$ in CaFeAsH. Figure 3(c) shows the Fermi surfaces of CaFeAsH and CaFeAsF. They each have two hole pockets (denoted as $\alpha_2$ and $\gamma$) and two electron pockets with the same orbital nature. The primary difference between them is the shape of the smallest hole pocket (denoted as $\alpha_1$) around the

Γ to Z line, which is indicated by the red line. A sandglass-shaped pocket exists in CaFeAsH, compared with a cylinder-shaped pocket in CaFeAsF. Figure 4 shows the Fermi surfaces of CaFe$_{1-x}$Co$_x$AsH and CaFe$_{1-x}$Co$_x$AsF, with $x = 0.10$ which corresponds to the optimal electron doping level. As described in the supplementary information (Figs. S1 and S2 of [29]), a rigid band model is valid for Co-substitution. That is, the band structure of CaFe$_{1-x}$Co$_x$As(H, F) is similar to that of CaFeAs(H, F), with the $E_F$ shift corresponding to the number of electron supplied from Co. The $E_F$ shift results in the smallest hole-pockets in both systems remaining largely unchanged by Co-doping. However, two hole pockets diminish and two electron pockets slightly enlarge with increasing Co content. This robustness of the α$_1$ hole pocket against Co-doping is attributed to the dispersion of the band producing the α$_1$ hole pocket being larger than that of the electron, α$_2$ or γ hole pocket.

**IV. DISCUSSION**

The difference in the electronic structures of CaFeAsH and CaFeAsF is first discussed. The differences in the calculation are the structural parameters ($a$, $c$, $z_{As}$ and $z_{Ca}$) and the anion species within the blocking layer (H$^-$ or F$^-$). However, the band structure calculated for CaFeAsF using the structural parameters for CaFeAsH does not reproduce the Fe 3$d$ band crossing the $E_F$ in the Γ-Z path (See Supplemental Material, Fig. S3 of [29]). Therefore, the three-dimensional (3D) nature originates from the contribution of the hydrogen 1$s$ state, despite it being located far from the

$E_F$. Figure 3(a) shows that the hydrogen 1$s$ states energetically overlap well with As 4$p$, and slightly overlap with the Ca components. The fluorine 2$p$ states do not energetically overlap with the Ca components. This overlap results in the formation of weak covalent bonding. The covalency of energetically overlapped H 1$s$ and As 4$p$ states is investigated next. Figure 5(a) shows the contribution of As 4$p$ and H 1$s$ to the band dispersion. In CaFeAsH, hybridization of As 4$p$ with the H 1$s$ orbital is apparent in some parts of the bands. This indicates the presence of covalent bonding between them, despite that they are separated by 334 pm. No such covalency is observed in CaFeAsF, because each orbital contribution is distinctly separated between F 2$p$ and As 4$p$ states.

The effect of this covalent bonding on the Fermi surface is discussed next. The change in the band structure of CaFeAsH near the $E_F$ is explained by the schematic orbital configuration shown in Fig. 5(b). The H 1$s$ orbital covalently bonds with the As 4$p_z$ orbital, the latter which is overlapped with the lobes of the Fe $3d_{x^2-y^2}$ and $3d_{z^2}$ orbitals. The H 1$s$ orbital weakly overlaps with Ca 4$s$/3$d$ orbitals. The As 4$p$ orbitals mediate the inter-layer bonding of the Fe $3d_{x^2-y^2}$ and $3d_{z^2}$ orbitals in adjacent FeAs layers. This results in the 3D electronic structure observed in Fig. 3(b). The dispersion of As 4$p_z$-derived bands along the Γ-Z direction depends on the dimensionality of the crystal structure.[30] For example, the band width along Γ-Z is ≈ 0 eV in $Ca_4As_2O_6Fe_2As_2$ with its thick blocking layer, while it is 0.4 and 2 eV in LaFeAsO and $BaFe_2As_2$, respectively. In $BaFe_2As_2$, the energy band with Fe $3d_{z^2}$ character crosses the $E_F$, forming the 3D hole pocket. The 3D electronic nature of BaFe$_2$As$_2$ is apparent from its electron transport properties.[31] In CaFeAsH, the 3D electronic nature is caused by bonding passing through the As-H-Ca bond. Covalent bonding between H and As is also reflected in the crystal structure of CaFeAsH: the shorter $d_{\text{Ca-As}}$ distance in CaFe$_{1-x}$Co$_x$AsH in Fig. 1(e) results from the decreased As-H separation. Electrons in the As $4p_z$ orbital are partly utilized in bond formation. Thus, the Fe-As bond in CaFeAsH weakens and lengthens relative to that in CaFeAsF, as shown in Fig. 1(g).

Spin fluctuations arising from the Fermi surface nesting between the hole and electron pockets are a plausible explanation for the mediation of superconductivity in iron pnictides.[32,33] The primary difference in the Fermi surfaces of CaFeAsH and CaFeAsF is the dimensionality of the α$_1$ hole pocket around the Γ-Z line, as indicated by red line. The former is 3D and the latter is two-dimensional (2D). Both hole pockets are smaller, and their curvatures are much larger than those of the electron pockets. This means the α$_1$ hole pocket does not contribute effectively to the nesting, *i.e.*, the development of spin fluctuation. The size and shape of the α$_1$ hole pocket remain largely unchanged up to $x = 0.1$, indicating that its contribution to superconductivity is small.

This hydrogen effect leads us to further consider its role in hydrogen-substituted *Ln*FeAsO. Hydrogen substitution effectively forms 3D Fermi surfaces, even for materials with 2D crystal structures. However, this change in Fermi surface does not affect their superconducting properties.

Hydrogen is an effective dopant for electron generation via oxygen site substitution, similarly to fluorine in $Ln$FeAsO. This indicates that incorporating hydrides enhances the 3D nature of 1111-type compounds, without suppressing their superconductivity.

## V. SUMMARY

Superconductivity was observed in CaFe$_{1-x}$Co$_x$H, and its properties were compared with those of CaFe$_{1-x}$Co$_x$AsF. The maximum $T_c$ and width of the superconducting dome of CaFe$_{1-x}$Co$_x$AsH are almost the same as those of CaFe$_{1-x}$Co$_x$AsF. The calculated electronic structure of CaFeAsH differs from that of CaFeAsF. The former has a 3D hole surface, with a highly 3D nature. This fact is caused by covalent bonding between energetically overlapped As $4p$ and H $1s$ bands. This 3D hole surface does not interfere with superconductivity because poor nesting between this small hole surface and electron surfaces causes the unfavorable development of excitations, such as spin and/or charge. Hydrogen incorporated within the blocking layer acts as an indirect electron dopant, without interfering with the superconductivity.

FIGURE CAPTIONS

FIG. 1. (Color online) Structural details of $CaFe_{1-x}Co_xAsH$. (a) Powder XRD pattern of $CaFe_{0.88}Co_{0.12}AsH$. Red and black traces indicate observed and Rietveld-fitted patterns, respectively. The differences between them (blue) and Bragg positions of the main phase (green) and Fe impurity (wine-red) are also shown. Reflections from unknown phases are denoted by orange asterisks. (b) Analyzed $x$ content as a function of $x_{nom}$. (c) and (d) Lattice parameters $a$ and $c$, respectively, as a function of $x$. (e), (f) and (g) Fe-As bond length, Ca-(H,F) bond length and distance between the FeAs and blocking layer ($d_{Ca-As}$), respectively, as a function of $x$.

FIG. 2. (Color online) Electronic and magnetic properties of $CaFe_{1-x}Co_xAsH$. (a) and (b) $\rho$-$T$ profiles for $x$ = 0.02–0.12 and 0.17–0.31, respectively. (c) and (d) Zero-field cooling (ZFC) $4\pi\chi$-$T$ curves measured on powdered samples under the magnetic field (H) of 10 Oe for x = 0.02-0.12 and 0.17-0.31. (e) SVF estimated from $M$-$H$ curves at 2 K and 10 Oe (red circle) and the volume fraction of Co-rich impurity (black line and triangle) (f) $x$-$T$ diagram of $CaFe_{1-x}Co_xAsH$ compared with date reported for $CaFe_{1-x}Co_xAsF$.[28]

FIG. 3. (Color online) Calculated electronic structures of CaFeAsH and CaFeAsF. (a) Total DOS and PDOS of CaFeAsH (left) and CaFeAsF (right). (b) Band structures along directions of high

symmetry in the Brillouin zone. Thick bands (blue) show the amounts of Fe-$d_{xy}$, $d_{yz/zx}$, $d_{x^2-y^2}$ and $d_{z^2}$ character. (c) Cross sections of Fermi surfaces in the $k_z = 0$ (top) and $k_x = k_y$ (bottom) planes.

FIG. 4. (Color online) Cross sections of Fermi surfaces of CaFe$_{0.9}$Co$_{0.1}$AsH and CaFe$_{0.9}$Co$_{0.1}$AsF, in the $k_z = 0$ (top) and $k_x = k_y$ (bottom) planes.

FIG. 5. (Color online) Contribution of arsenic, hydrogen and fluorine atomic orbitals to the electronic structures of CaFeAsH and CaFeAsF. (a) Thickness of bands shows the amounts of As-$p$ (green), H-$s$ (red) and F-$2p$ (purple) character. (b) and (c) Schematics showing the configurations of the H-1s, As-4$p$, Fe-3$d_{x^2-y^2}$ and -3$d_{z^2}$ orbitals.

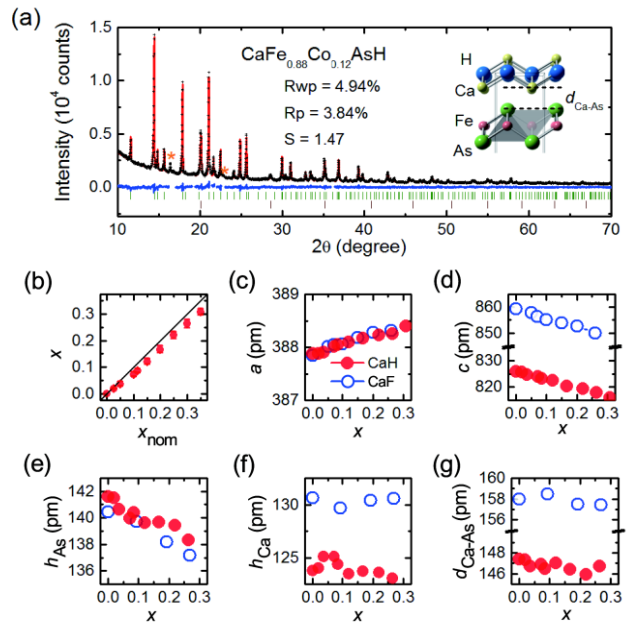

FIG. 1

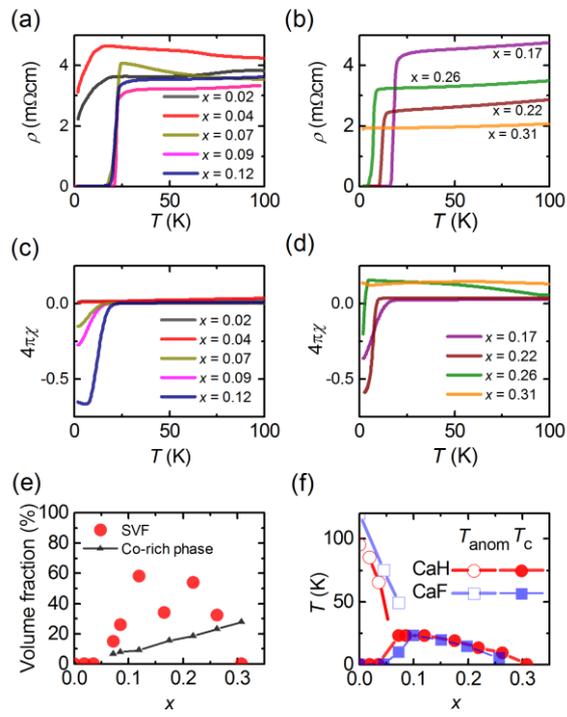

FIG. 2

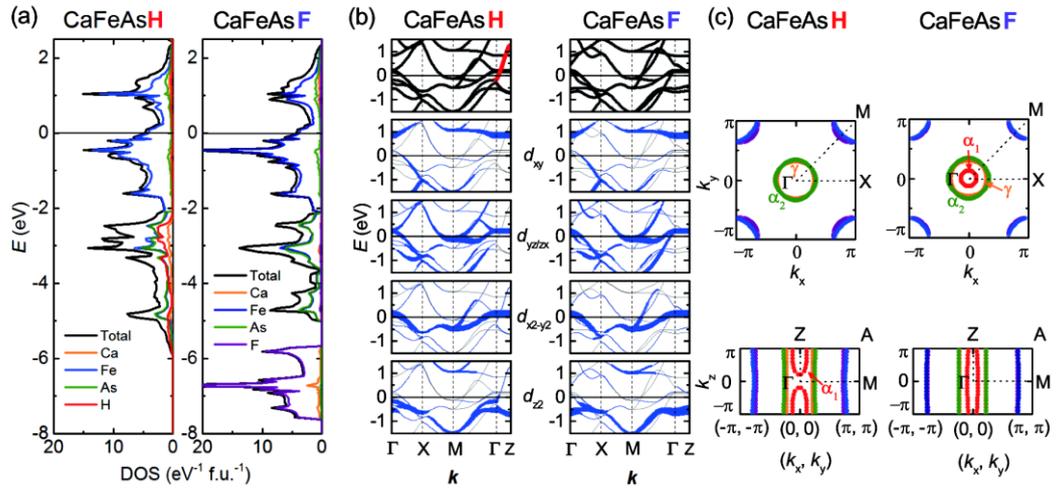

FIG. 3

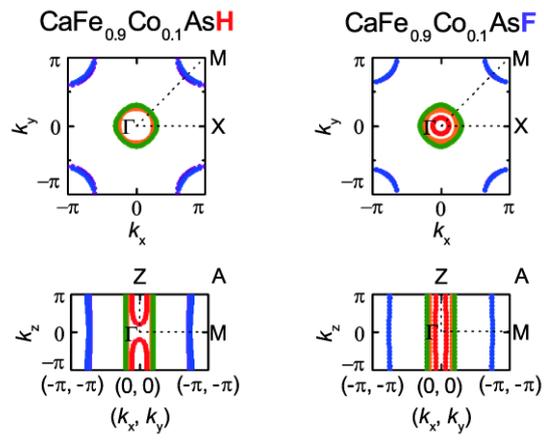

FIG. 4

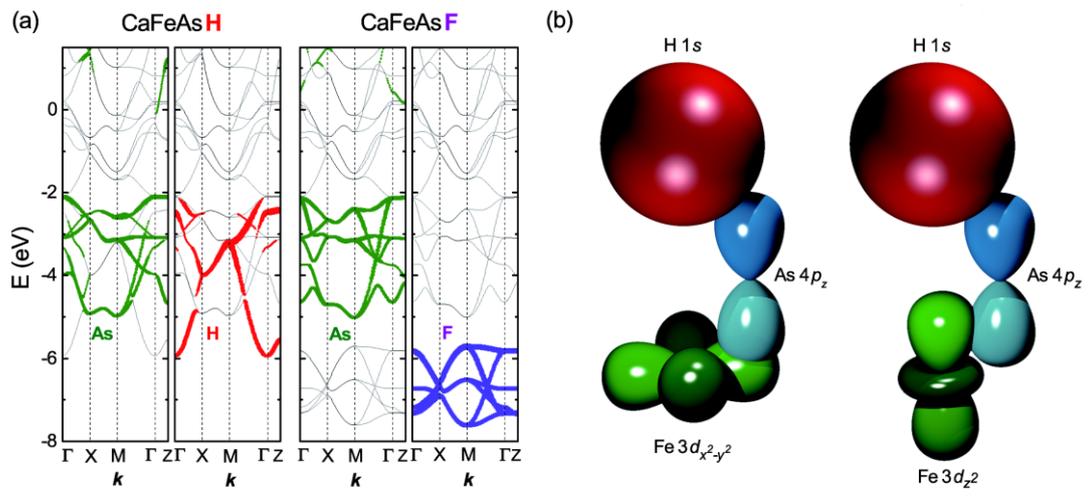

FIG. 5